\documentclass[preprints,article,accept,moreauthors,pdftex]{Definitions/mdpi}

\firstpage{1} 
\pubvolume{1}
\issuenum{1}
\articlenumber{0}
\pubyear{2022}
\copyrightyear{2022}
\externaleditor{Academic Editor: Robert E. Wilson; Walter Van Hamme} 
\datereceived{22 July 2022} 
\dateaccepted{14 September 2022} 
\datepublished{} 
\hreflink{https://doi.org/} 
\pdfoutput=1

\Title{Fundamental Properties of Late-Type Stars in Eclipsing Binaries}

\TitleCitation{Fundamental Properties of Late-Type Stars in Eclipsing Binaries}

\Author{Juan Carlos Morales $^{1,2,}$, Ignasi Ribas $^{1,2}$, \'Alvaro Gim\'enez $^{3}$ and  David Baroch $^{1,2}$}

\AuthorNames{Juan Carlos Morales, Ignasi Ribas, \'Alvaro Gim\'enez, and David Baroch }

\AuthorCitation{Morales, J.C.; Ribas, I.; Gim\'enez, \'A.; Baroch, D. }

\address{%
$^{1}$ \quad Institut de Ci\`encies de l'Espai (ICE, CSIC), Campus UAB, c/ Can Magrans s/n, \mbox{E-08193 Bellaterra, Barcelona, Spain}\\
$^{2}$ \quad Institut d'Estudis Espacials de Catalunya (IEEC), c/ Gran Capit\`a 2-4, E-08034 Barcelona, Spain\\
$^{3}$ \quad Centro de Astrobiolog\'{\i}a (CSIC-INTA), Ctra. Ajalvir, km 4., E-28850 Torrej\'on de Ardoz, Madrid, Spain
}

\corres{Correspondence: morales@ice.cat}

\abstract{Evidence from the analysis of eclipsing binary systems revealed that late-type stars are larger and cooler than predicted by models, and that this is probably caused by stellar magnetic activity. In this work, we revisit this problem taking into account the advancements in the last decade. We provide and updated a list of 32 eclipsing binary or multiple systems, including at least one star with a mass $\lesssim 0.7$\,M$_{\odot}$ and with mass and radius measured to an accuracy better than 3\%. The~comparison with stellar structure and evolution theoretical models reveals an overall discrepancy of about 7\% and $-$4\% for the radius and effective temperature, respectively, and that it may be larger than previously found below the full convection boundary.  Furthermore, the hypothesis of stellar activity is reinforced by the comparison of different systems with similar components. Further eclipsing binaries with accurately determined masses and radii, and with estimated activity levels, as well as the implementation of magnetic activity in theoretical models will help to improve our knowledge of low-mass stars, which are prime targets for exoplanet surveys.}


\begin{document}

\section{Introduction}
\label{sec_introduction}
Several astronomical surveys aim at late-type stars because they are excellent targets to look for Earth-like planets within the habitable zone. Due to their small mass, the~radial velocity signature imprinted by planets around them are larger, and~the habitable zone is closer, therefore their transit probability is also larger (see e.g.,~\citep{Marcy1998,Bonfils2013,Quirrenbach2018}). However, their structure is not fully understood yet due to the difficulty to accurately measure their fundamental properties, from~which those of the exoplanet they host~depend.

Double-lined eclipsing binary systems (hereafter, DLEBs) were revealed to be a unique opportunity to derive the masses and radii of stars in a fundamental way, almost independent of any assumptions. This is achieved by analyzing both the photometric light curves showing the mutual eclipses of the components, which provide the relative size of the stars and the orbital inclination, and~the radial velocities, which provide the minimum masses and absolute dimensions of the orbit. Such studies can yield the masses and radii of the component stars with a precision better than 3\% that allows us to thoroughly test the predictions of stellar models (see e.g.,~\citep{Andersen1991,Torres2010,Serenelli2021}). Furthermore, due to their accuracy, the~masses and radii of DLEBs are also used to calibrate empirical mass--radius relationships from which the properties of single stars are then computed (e.g., \citep{Schweitzer2019,Mann2019}).

However, several studies of DLEBs composed of late-K and M dwarfs found discrepancies between the results of the analysis of observations and the theoretical structure and evolution models. The~general conclusion was that the observed radius of low-mass stars in eclipsing binaries is $\sim$5--10\% larger than predicted by models, while effective temperatures are $\sim$10\% cooler (see e.g.,~\citep{Popper1986,Torres2002,Ribas2003,Ribas2006}). Interestingly, the~luminosities are in agreement. Stellar magnetic activity was suggested as the probable cause of such discrepancies \citep{Torres2002,LopezMorales2005,LopezMorales2007a}. Known DLEBs with low-mass components are typically close systems with orbital periods of few days, with~components tidally locked and rotating synchronously \citep{Mazeh2008}. This fast rotation induces intense magnetic fields, which generate cold spots in the surface of the stars. Studies on theoretical models found that the presence of such spots significantly affects the structure of stars \citep{Mullan2001,Chabrier2007,Feiden2012,Feiden2013,Feiden2014a} and models including such effects have been recently published \citep{Somers2020}. Moreover, some studies found that stellar spots can also cause some bias on the determination of the fundamental properties from light curve analyses \citep{Morales2010,Windmiller2010}.

In this paper, we compile a list of the up-to-date known late-type stars in eclipsing binary and multiple systems with well-determined masses and radii with uncertainties below the 3\% level to revisit the comparison between the observations and updated low-mass stellar models. We discuss the present state of this long-standing issue and provide some ideas for future~studies.

\section{Sample of Low-Mass~Systems}
\label{sec:sample}

Torres~et~al.~\cite{Torres2010} compiled a list of stars in DLEBs with reliable analyses providing the masses and radii of the components with uncertainties below the 3\% limit. Only four systems with late-K and M dwarf components were included in this list, namely the well-known classical cases of GU Boo \citep{LopezMorales2005}, YY Gem \citep{Torres2002}, CU Cnc \citep{Ribas2003}, and CM Dra \citep{Morales2009a}. In~the recent years, the~large number of ground and space-based surveys to look for exoplanets yielded also several eclipsing binary systems as a byproduct. In~Table~\ref{tab:sample}, we list the properties of low-mass stars with accurately measured masses and radii. We have limited our sample to DLEBs with at least one component with a mass below $\sim$0.7 M$_{\odot}$ to focus our attention on the late-K and M dwarf domain, which is particularly interesting nowadays for exoplanet surveys. In~order to perform a meaningful comparison with stellar structure models, we also restricted the sample to those systems with main sequence components and mass and radius uncertainties $\leq$3\%. Only the reported uncertainties in the publications were considered, we did not apply any further selection based on the published analysis. This resulted in 28 DLEBs.

The extremely precise and continuous monitoring of space telescopes also allows to measure the masses and radii of the components \textls[-15]{in multiple systems showing mutual eclipses. This has already been achieved either for triple stellar systems such as KOI-126~\citep{Carter2011},} or~binary systems with circumbinary planets such as Kepler-16 \citep{Doyle2011}, Kepler-453~\citep{Welsh2015}, and~Kepler-47 \citep{Orosz2019}. The~timing of the eclipses and transits between the different components becomes a very useful tool to accurately compute their masses without the necessity of spectroscopic orbits. On~the other hand, the~radius is determined from the eclipses and transit shapes as usual, with~the additional constraint of the dynamical effects of the multiple systems. Those systems are also included in our sample.

\textls[-15]{The 32 systems listed in Table~\ref{tab:sample} contain a total of 52 late-type dwarfs with \mbox{$M \lesssim 0.7$\,M$_{\odot}$}} with fundamental properties determined to be a precision better than 3\%. Most of the targets have short orbital periods; thus, their orbital and rotation periods are synchronized; thus, they are fast rotators and hence magnetically active. Indeed, studies about the tidal evolution of binary stars demonstrate that close binaries are synchronized and circularized relatively fast within a few million  years; thereby, systems with orbital periods below $\sim$20 days may be synchronized by $\sim$1 Ga (see e.g.,~\citep{Mazeh2008,Claret1995,Claret1997,Barker2022}). Hints of stellar activity are reported for all of the systems, except for EBLM J0113+31 \citep{Maxted2022}. The~most common signature is out-of-eclipse variability due to spots of a few percent in flux; however,~other activity indicators such as chromospheric emission in the H$\alpha$ and Ca\,H\&K lines, enhanced X-ray emission, or~flaring activity are reported for several systems (e.g., \citep{Torres2002,Ribas2006,Morales2009a}). As~mentioned, stellar activity may be driven by the fast rotation of the components, which is typically synchronized with the orbital motion for short period systems. For~this reason, we included in Table~\ref{tab:sample} the rotation velocity ($v_{\rm sync}$) of the components as computed from the radius of each star, and~the orbital period and eccentricity of each system assuming periastron pseudo-synchronization~\cite{Hut1983}. Interestingly, large photometric surveys are also providing long-period systems that in principle should be less prone to be magnetically active. Actually, rotation rates slightly faster and significantly slower than expected from synchronization are reported for the long period systems Kepler-453 \citep{Welsh2015} and LSPM J1112+7626 \citep{Irwin2011}, respectively, which may indicate that they have not yet reached spin-orbit synchronization.

\begin{table}[H]
\tablesize{\tiny}
\caption{List of DLEBs and multiple systems with a least one stellar component with a mass $\lesssim 0.7$\,M$_{\odot}$ and mass and radius uncertainties below 3\%.}
\label{tab:sample}
\begin{adjustwidth}{-\extralength}{0cm}
		\newcolumntype{C}{>{\centering\arraybackslash}X}
\newcolumntype{L}{>{\raggedright\arraybackslash}X}
		\begin{tabularx}{\fulllength}{LcCCCCCCc}
\toprule 
\textbf{Name} & \textbf{Com.} & \boldmath{$P$} & \boldmath{$M$} & \boldmath{$R$} & \boldmath{$T_{\rm eff}$} & \boldmath{$v_{\rm sync}$}  & \textbf{[Fe/H]}  & \textbf{Ref.} \\
 & & \textbf{[d]} & \boldmath{[M$_{\odot}$]} & \boldmath{[R$_{\odot}$]} & \textbf{[K]} & \boldmath{\textbf{[km\,s}$^{-1}$]} & \textbf{[dex]}  &  \\
\midrule
Eclipsing binary systems  & & & & & & & & \\
\midrule
 \multirow{2}{*}{NGTS J052218.2-250710.4} & A &  \multirow{2}{*}{1.7477} &0.17391 $\pm$ 0.00126 & 0.2045 $\pm$ 0.0048 & 2995 $\pm$ 95 & 5.93 $\pm$ 0.13 & \multirow{2}{*}{$\cdots$} & \multirow{2}{*}{\cite{Casewell2018}} \\
& B & &0.17418 $\pm$ 0.00126 & 0.2168 $\pm$ 0.00475 & 2997 $\pm$ 84 & 6.29 $\pm$ 0.14 &  \\
\multirow{2}{*}{CM Dra} & A &  \multirow{2}{*}{1.2684} &0.23102 $\pm$ 0.00089 & 0.2534 $\pm$ 0.0019 & 3130 $\pm$ 70 & 10.22 $\pm$ 0.08 & \multirow{2}{*}{$-$0.3} &\multirow{2}{*}{\cite{Torres2010}} \\
& B & &0.2141 $\pm$ 0.0008 & 0.2398 $\pm$ 0.0018 & 3120 $\pm$ 70 & 9.67 $\pm$ 0.08 &  \\
\multirow{2}{*}{LP 661-13} & A &  \multirow{2}{*}{4.7044} &0.30795 $\pm$ 0.00084 & 0.3226 $\pm$ 0.0033 & $\cdots$ &3.47 $\pm$ 0.04 & \multirow{2}{*}{$-$0.07 $\pm$ 0.07} &\multirow{2}{*}{\cite{Dittmann2017}} \\
& B & &0.194 $\pm$ 0.00034 & 0.2174 $\pm$ 0.0023 & $\cdots$ &2.338 $\pm$ 0.024 &  \\
\multirow{2}{*}{LSPMJ1112+7626} & A &  \multirow{2}{*}{41.0324} &0.3951 $\pm$ 0.0022 & 0.3815 $\pm$ 0.003 & 3130 $\pm$ 165 & 0.789 $\pm$ 0.006 & \multirow{2}{*}{$\cdots$} & \multirow{2}{*}{\citep{Irwin2011}} \\
& B & &0.2749 $\pm$ 0.0011 & 0.2999 $\pm$ 0.0044 & 3015 $\pm$ 166 & 0.620 $\pm$ 0.009 &  \\
\multirow{2}{*}{NGTS 0002-29} & A &  \multirow{2}{*}{1.098} &0.3978 $\pm$ 0.0033 & 0.4037 $\pm$ 0.0048 & 3372 $\pm$ 40 & 18.62 $\pm$ 0.22 & \multirow{2}{*}{0.04 $\pm$ 0.04} &\multirow{2}{*}{\cite{Smith2021}} \\
& B & &0.2245 $\pm$ 0.0018 & 0.2759 $\pm$ 0.0055 & 3231 $\pm$ 34 & 12.72 $\pm$ 0.24 &  \\
\multirow{2}{*}{CU Cnc} & A &  \multirow{2}{*}{2.7715} &0.4349 $\pm$ 0.0012 & 0.4323 $\pm$ 0.0055 & 3160 $\pm$ 150 & 7.89 $\pm$ 0.10 & \multirow{2}{*}{0.0} &\multirow{2}{*}{\cite{Torres2010}} \\
& B & &0.3992 $\pm$ 0.0009 & 0.3916 $\pm$ 0.0094 & 3125 $\pm$ 150 & 7.15 $\pm$ 0.17 &  \\
\multirow{2}{*}{HAT-TR-318-007} & A &  \multirow{2}{*}{3.344} &0.448 $\pm$ 0.011 & 0.4548 $\pm$ 0.0036 & 3190 $\pm$ 100 & 7.07 $\pm$ 0.06 & \multirow{2}{*}{0.298 $\pm$ 0.08} &\multirow{2}{*}{\cite{Hartman18}} \\
& B & &0.2721 $\pm$ 0.0042 & 0.2913 $\pm$ 0.0024 & 3100 $\pm$ 100 & 4.53 $\pm$ 0.04 &  \\
\multirow{2}{*}{MG1-2056316} & A &  \multirow{2}{*}{1.7228} &0.469 $\pm$ 0.002 & 0.441 $\pm$ 0.002 & 3460 $\pm$ 180 & 12.95 $\pm$ 0.06 & \multirow{2}{*}{$\cdots$} & \multirow{2}{*}{\cite{Kraus2011a}} \\
& B & &0.382 $\pm$ 0.001 & 0.374 $\pm$ 0.002 & 3320 $\pm$ 180 & 10.98 $\pm$ 0.06 &  \\
\multirow{2}{*}{MG1-646680} & A &  \multirow{2}{*}{1.6375} &0.499 $\pm$ 0.002 & 0.457 $\pm$ 0.006 & 3730 $\pm$ 20 & 14.12 $\pm$ 0.18 & \multirow{2}{*}{$\cdots$} & \multirow{2}{*}{\cite{Kraus2011a}} \\
& B & &0.443 $\pm$ 0.002 & 0.427 $\pm$ 0.006 & 3630 $\pm$ 20 & 13.19 $\pm$ 0.18 &  \\
\multirow{2}{*}{MG1-78457} & A &  \multirow{2}{*}{1.5862} &0.527 $\pm$ 0.002 & 0.505 $\pm$ 0.008 & 3330 $\pm$ 60 & 16.11 $\pm$ 0.25 & \multirow{2}{*}{$-$1.55 $\pm$ 0.05} &\multirow{2}{*}{\cite{Kraus2011a}} \\
& B & &0.491 $\pm$ 0.001 & 0.471 $\pm$ 0.007 & 3270 $\pm$ 60 & 15.02 $\pm$ 0.22 &  \\
\multirow{2}{*}{NSVS01031772} & A &  \multirow{2}{*}{0.3681} &0.53 $\pm$ 0.014 & 0.559 $\pm$ 0.014 & 3750 $\pm$ 150 & 76.8 $\pm$ 1.9 & \multirow{2}{*}{$\cdots$} & \multirow{2}{*}{\cite{LopezMorales2007b}} \\
& B & &0.514 $\pm$ 0.013 & 0.518 $\pm$ 0.013 & 3600 $\pm$ 150 & 71.2 $\pm$ 1.8 &  \\
\multirow{2}{*}{MG1-116309} & A &  \multirow{2}{*}{0.8271} &0.567 $\pm$ 0.002 & 0.552 $\pm$ 0.013 & 3920 $\pm$ 80 & 33.76 $\pm$ 0.76 & \multirow{2}{*}{$-$1.19 $\pm$ 0.04} &\multirow{2}{*}{\cite{Kraus2011a}} \\
& B & &0.532 $\pm$ 0.002 & 0.532 $\pm$ 0.008 & 3810 $\pm$ 80 & 32.54 $\pm$ 0.50 &  \\
\multirow{2}{*}{MG1-506664} & A &  \multirow{2}{*}{1.5484} &0.584 $\pm$ 0.002 & 0.56 $\pm$ 0.005 & 3730 $\pm$ 90 & 18.30 $\pm$ 0.16 & \multirow{2}{*}{$\cdots$} & \multirow{2}{*}{\cite{Kraus2011a}} \\
& B & &0.544 $\pm$ 0.002 & 0.513 $\pm$ 0.008 & 3610 $\pm$ 90 & 16.76 $\pm$ 0.26 &  \\
YY Gem & A\&B &  0.8143 &0.5992 $\pm$ 0.0047 & 0.6194 $\pm$ 0.0057 & 3820 $\pm$ 100 & 38.48 $\pm$ 0.36 & 0.0 & \cite{Torres2010} \\
\multirow{2}{*}{GU Boo} & A &  \multirow{2}{*}{0.4887} &0.6101 $\pm$ 0.0064 & 0.627 $\pm$ 0.016 & 3920 $\pm$ 130 & 64.9 $\pm$ 1.7 & \multirow{2}{*}{$\cdots$} & \multirow{2}{*}{\cite{Torres2010}} \\
& B & &0.5995 $\pm$ 0.0064 & 0.624 $\pm$ 0.016 & 3810 $\pm$ 130 & 64.6 $\pm$ 1.7 &  \\
\multirow{2}{*}{HIP 41431} & A &  \multirow{2}{*}{2.9300} &0.625 $\pm$ 0.010 & 0.588 $\pm$ 0.012 & 4043 $\pm$ 60 & 10.55 $\pm$ 0.22 & \multirow{2}{*}{$\cdots$} & \multirow{2}{*}{\cite{Borkovits2019}} \\
& B & &0.614 $\pm$ 0.012 & 0.576 $\pm$ 0.012 & 3986 $\pm$ 60 & 10.33 $\pm$ 0.22 &  \\
\multirow{2}{*}{KIC 9821078} & A &  \multirow{2}{*}{8.4294} &0.67 $\pm$ 0.01 & 0.662 $\pm$ 0.001 & $\cdots$ &4.233 $\pm$ 0.007 & \multirow{2}{*}{$\cdots$} & \multirow{2}{*}{\cite{Han2019}} \\
& B & &0.52 $\pm$ 0.01 & 0.478 $\pm$ 0.001 & $\cdots$ &3.056 $\pm$ 0.007 &  \\
\multirow{2}{*}{BD-15 2429} & A &  \multirow{2}{*}{1.5285} &0.7029 $\pm$ 0.0045 & 0.694 $\pm$ 0.011 & 4230 $\pm$ 200 & 23.20 $\pm$ 0.35 & \multirow{2}{*}{$\cdots$} & \multirow{2}{*}{\cite{Helminiak2011a}} \\
& B & &0.6872 $\pm$ 0.0049 & 0.699 $\pm$ 0.014 & 4080 $\pm$ 200 & 23.37 $\pm$ 0.46 &  \\
\multirow{2}{*}{M55 V54} & A &  \multirow{2}{*}{9.2692} &0.726 $\pm$ 0.015 & 1.006 $\pm$ 0.009 & 6246 $\pm$ 71 & 7.24 $\pm$ 0.06 & \multirow{2}{*}{$-$1.86 $\pm$ 0.15} &\multirow{2}{*}{\cite{Kaluzny2014}} \\
& B & &0.555 $\pm$ 0.008 & 0.528 $\pm$ 0.005 & 5020 $\pm$ 95 & 3.80 $\pm$ 0.04 &  \\
\multirow{2}{*}{RXJ0239.1-1028} & A &  \multirow{2}{*}{2.0719} &0.73 $\pm$ 0.009 & 0.741 $\pm$ 0.004 & 4645 $\pm$ 20 & 18.09 $\pm$ 0.10 & \multirow{2}{*}{$\cdots$} & \multirow{2}{*}{\cite{LopezMorales2007b}} \\
& B & &0.693 $\pm$ 0.006 & 0.703 $\pm$ 0.002 & 4275 $\pm$ 15 & 17.17 $\pm$ 0.05 &  \\
\multirow{2}{*}{NGC2204-S892} & A &  \multirow{2}{*}{0.4518} &0.733 $\pm$ 0.005 & 0.719 $\pm$ 0.014 & 4200 $\pm$ 100 & 80.5 $\pm$ 1.6 & \multirow{2}{*}{$\cdots$} & \multirow{2}{*}{\cite{Rozyczka2009}} \\
& B & &0.662 $\pm$ 0.005 & 0.68 $\pm$ 0.017 & 3940 $\pm$ 110 & 76.2 $\pm$ 1.9 &  \\
\multirow{2}{*}{UCAC3 127-192903} & A &  \multirow{2}{*}{2.293} &0.8035 $\pm$ 0.0086 & 1.147 $\pm$ 0.01 & 6088 $\pm$ 108 & 25.31 $\pm$ 0.22 & \multirow{2}{*}{$-$1.18 $\pm$ 0.02} &\multirow{2}{*}{\cite{Kaluzny2013}} \\
& B & &0.605 $\pm$ 0.0044 & 0.611 $\pm$ 0.0092 & 4812 $\pm$ 125 & 13.48 $\pm$ 0.21 &  \\
\multirow{2}{*}{KIC 6131659} & A &  \multirow{2}{*}{17.5278} &0.922 $\pm$ 0.007 & 0.88 $\pm$ 0.0028 & 5789 $\pm$ 50 & 2.540 $\pm$ 0.008 & \multirow{2}{*}{$-$0.23 $\pm$ 0.2} &\multirow{2}{*}{\cite{Bass2012}} \\
& B & &0.685 $\pm$ 0.005 & 0.6395 $\pm$ 0.0061 & 4609 $\pm$ 32 & 1.846 $\pm$ 0.018 &  \\
\multirow{2}{*}{EPIC 247605441} & A &  \multirow{2}{*}{1.6534} &0.934 $\pm$ 0.017 & 1.058 $\pm$ 0.023 & 5668 $\pm$ 71 & 32.3 $\pm$ 0.7 & \multirow{2}{*}{$-$0.26 $\pm$ 0.26} &\multirow{2}{*}{\cite{Helminiak2021}} \\
& B & &0.409 $\pm$ 0.005 & 0.408 $\pm$ 0.009 & 3590 $\pm$ 100 & 12.48 $\pm$ 0.28 &  \\
\multirow{2}{*}{ASAS J065134-2111.5} & A &  \multirow{2}{*}{8.2196} &0.956 $\pm$ 0.012 & 0.997 $\pm$ 0.004 & 5500 $\pm$ 100 & 6.277 $\pm$ 0.024 & \multirow{2}{*}{0.09 $\pm$ 0.13} &\multirow{2}{*}{\cite{Helminiak2019}} \\
& B & &0.674 $\pm$ 0.005 & 0.69 $\pm$ 0.007 & 3970 $\pm$ 110 & 4.34 $\pm$ 0.04 &  \\
\multirow{2}{*}{IM Vir} & A &  \multirow{2}{*}{1.3086} &0.981 $\pm$ 0.012 & 1.061 $\pm$ 0.016 & 5570 $\pm$ 100 & 41.01962 $\pm$ 0.64 & \multirow{2}{*}{$-$0.3} &\multirow{2}{*}{\cite{Morales2009b}} \\
& B & &0.6644 $\pm$ 0.0048 & 0.681 $\pm$ 0.013 & 4250 $\pm$ 130 & 26.33 $\pm$ 0.51 &  \\
\multirow{2}{*}{V530 Ori} & A &  \multirow{2}{*}{6.1108} &1.0038 $\pm$ 0.0066 & 0.98 $\pm$ 0.013 & 5890 $\pm$ 100 & 9.68 $\pm$ 0.13 & \multirow{2}{*}{$-$0.12 $\pm$ 0.08} &\multirow{2}{*}{\cite{Torres2014}} \\
& B & &0.5955 $\pm$ 0.0022 & 0.5873 $\pm$ 0.0067 & 3880 $\pm$ 120 & 5.80 $\pm$ 0.06 &  \\
\multirow{2}{*}{EBLM J0113+31} & A &  \multirow{2}{*}{14.2768} &1.029 $\pm$ 0.025 & 1.417 $\pm$ 0.014 & 3.787 $\pm$ 0.003 & 10.00 $\pm$ 0.10 & \multirow{2}{*}{$-$0.3 $\pm$ 0.1} &\multirow{2}{*}{\cite{Maxted2022}} \\
 & B & &0.197 $\pm$ 0.003 & 0.215 $\pm$ 0.002 & 3.528 $\pm$ 0.005 & 1.517 $\pm$ 0.014 &   \\

\midrule
Eclipsing triple systems & & & & & & & & \\
\midrule
\multirow{3}{*}{KOI-126} & A & 33.9214 & 1.347 $\pm$ 0.032 & 2.20254 $\pm$ 0.0098 & 5875 $\pm$ 100 & $\cdots$ & \multirow{3}{*}{0.15 $\pm$ 0.08} & \multirow{3}{*}{\cite{Carter2011}}\\
& Ba &  \multirow{2}{*}{1.7671} &0.2413 $\pm$ 0.003 & 0.2543 $\pm$ 0.0014 & $\cdots$ & 7.62 $\pm$ 0.04 &  & \\
& Bb & &0.2127 $\pm$ 0.0026 & 0.2318 $\pm$ 0.0013 & $\cdots$ & 6.94 $\pm$ 0.04 & & \\
\midrule
Circumbinary planets & & & & & & & & \\
\midrule

\multirow{2}{*}{Kepler 16} & A &  \multirow{2}{*}{41.0792} &0.6897 $\pm$ 0.0035 & 0.6489 $\pm$ 0.0013 & 4450 $\pm$ 150 & 1.1166 $\pm$ 0.0022 & \multirow{2}{*}{$-$0.3 $\pm$ 0.2} &\multirow{2}{*}{\cite{Doyle2011}} \\
& B & &0.20255 $\pm$ 0.00066 & 0.22623 $\pm$ 0.00059 & $\cdots$ &0.3893 $\pm$ 0.0010 &  \\
\multirow{2}{*}{Kepler-453} & A &  \multirow{2}{*}{27.322} &0.944 $\pm$ 0.01 & 0.833 $\pm$ 0.011 & 5527 $\pm$ 100 & 1.715 $\pm$ 0.022 & \multirow{2}{*}{0.09 $\pm$ 0.1} &\multirow{2}{*}{\cite{Welsh2015}} \\
& B & &0.1951 $\pm$ 0.002 & 0.215 $\pm$ 0.0014 & 3226 $\pm$ 100 & 0.4428 $\pm$ 0.0029 &  \\
\multirow{2}{*}{Kepler-47} & A &  \multirow{2}{*}{7.4484} &0.957 $\pm$ 0.014 & 0.936 $\pm$ 0.005 & 5636 $\pm$ 100 & 6.738 $\pm$ 0.037 & \multirow{2}{*}{$-$0.25 $\pm$ 0.08} &\multirow{2}{*}{\cite{Orosz2019}} \\
& B & &0.342 $\pm$ 0.003 & 0.338 $\pm$ 0.002 & 3357 $\pm$ 100 & 2.433 $\pm$ 0.015 &  \\
\bottomrule
\end{tabularx}
	\end{adjustwidth}
\end{table}

\section{Models vs.~Observations}
\label{sec:modelvsobs}

The standard theoretical models of stellar structure and evolution typically do not include the effect of intense magnetic fields, or~the appearance of surface spots, which are associated to stellar activity. However, several studies analyzed their effect in the past years. For~instance, Mullan and MacDonald \citep{Mullan2001,MacDonald2012} introduced the strength of the magnetic field in stellar structure models as an additional parameter changing the criterion of the onset of convection. Strong magnetic fields reduce the efficiency of convection and change the structure of the star, producing stars with a larger radius and cooler effective temperatures in line with DLEBs. In~contrast, luminosities are also~changed.

Chabrier~et~al.~\cite{Chabrier2007} also studied the impact of magnetic activity in the Lyon stellar structure models~\cite{Baraffe1995}. They suggested that the inhibition of convection could be modeled changing the mixing length parameter ($\alpha$) of the convection. Moreover, they also introduced a second parameter ($\beta$) that accounts for the effect of spots in the photosphere. They~demonstrated that the presence of spots, which block the outgoing flux, does significantly change the structure of low-mass stars over all the mass range. On~the other hand, the~inhibition of convection is only important above the fully convective boundary. A~subsequent comparison of these models with well-known DLEBs revealed that they would reproduce the radius of the components when a spot coverage of about 35\% of the photosphere is assumed ($\beta=0.17$) after removing a possible $\sim$3\% systematic bias on the radii determined from light curves due to the variability caused by stellar spots \citep{Morales2010}. Magnetic activity effects were also inspected using the Dartmouth stellar models \citep{Feiden2012} by Feiden and Chaboyer \citep{Feiden2013,Feiden2014a}. They concluded that magnetic fields of few kG can explain the inflated radii of partially convective stars \citep{Feiden2013}, but~not for fully convective stars~\cite{Feiden2014a}. The~conclusions of all these works confirm the hypothesis that stellar activity and magnetic fields play a key role in the stellar structure of low-mass stars, also from the theoretical point of~view.

More recently, a~new set of stellar models, dubbed SPOTS~\cite{Somers2020}, including the presence of stellar spots have been published. A~parameter accounting for the filling factor of spots  is implemented in these models~\cite{Somers2015}. The~surface \textls[-15]{inhomogeneities block the outgoing flux of the stars so their structure is altered. They reproduce the results of the work by Chabrier~et~al.~\cite{Chabrier2007}. The~filling factor ($f$) is computed assuming a phenomenological effective temperature difference} between the spots and the photosphere~\cite{Berdyugina2005}. This means that the actual fraction of photoshpere covered by spots may be different to $f$; it depends on the real temperature contrast of the surface features with respect to the~photosphere.

\begin{figure}[H]
\includegraphics[width=12 cm]{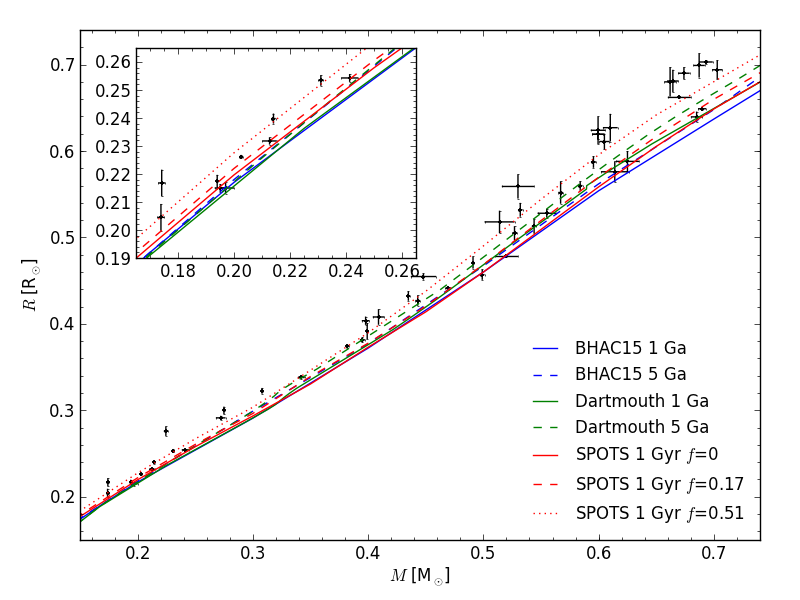}
\includegraphics[width=12 cm]{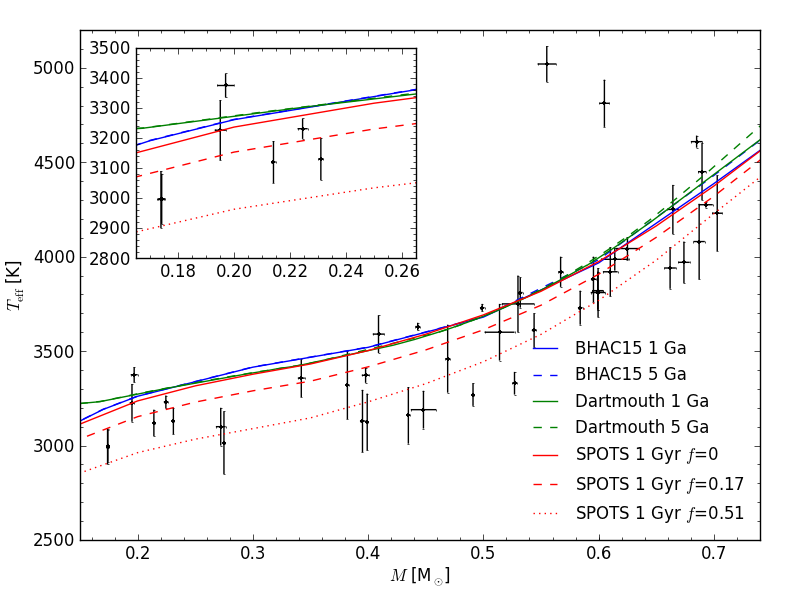}
\caption{$M-R$ (\textbf{top}) and $M-T_{\rm eff}$ (\textbf{bottom}) relationships for the late-type stars listed in Table\,\ref{tab:sample}. BHAC15, Darmouth, and SPOTS stellar models with different parameters are plotted as labeled. Inset plots show a zoom in into the lower mass~domain. \label{fig:plot_M-R-Teff}}
\end{figure}

Figure \ref{fig:plot_M-R-Teff} shows the mass--radius ($M-R$) and mass--effective temperature ($M-T_{\rm eff}$) relationships for low-mass stars from theoretical models compared with the measured values of the stars with $M \lesssim 0.7$\,M$_{\odot}$ in our sample of eclipsing and circumbinary systems in Table\,\ref{tab:sample}. Here, we compare the observations only with models publicly available: the more recent version of the Lyon stellar structure models \footnote{\url{http://perso.ens-lyon.fr/isabelle.baraffe/BHAC15dir/}}, with~updated molecular lines and newly calibrated convection parameters (hereafter BHAC15,~\cite{Baraffe2015}), the~Dartmouth stellar models\footnote{\url{http://stellar.dartmouth.edu/models/}} \citep{Dotter2008} for 1 and 5 Ga, and~the 1 Ga SPOTS models\footnote{\url{https://doi.org/10.5281/zenodo.3593339}} with different values of the filling factor parameter~\cite{Somers2020}. Differences between these sets of stellar models not including the effect of photospheric spots are small except for the larger masses, where the modeling of the convective layer of the star plays a significant role on the stellar structure (see e.g.,~\citep{Chabrier2007}). Taking as reference the 1\,Ga BHAC15 isochrone, the~mean radii differences between observations and models are 6.6\% (standard deviation, $\sigma=4.2$\%) for the 52 low-mass stars with available radii. In~the case of the effective temperature, the~mean difference with respect to the models is $-$2.5\% ($\sigma=7.6$\%) for the 45 stars
with measured temperatures; however, the~determination of the absolute effective temperature from spectrophotometric data is less constrained and depends on the determination of the absolute temperature of one of the components and the temperature ratio derived from multi-band light curves of the eclipsing binary systems; therefore, its comparison is not as straightforward as for the radii. Actually, our computed mean $T_{\rm eff}$ discrepancy is significantly influenced by the low-mass components of M55 V54 \citep{Kaluzny2014} and UCAC3 127-192903 \citep{Kaluzny2013}, which are suggested to be members of the globular clusters M55 and M4, respectively. The~proposed old age and poor metallicity of the systems, and~possible biases in the estimation of the primary $T_{\rm eff}$ from $B-V$ colors may be responsible for the large difference with respect to theoretical stellar models (see \citep{Kaluzny2013} for further details). Removing these systems from the sample, the~mean temperature difference is $-$3.9\% ($\sigma=4.4$\%).

For the sample of stars in Table~\ref{tab:sample}, the~radii discrepancy between fully convective stars, those with a mass $\lesssim 0.35$\,M$_{\odot}$, and~partially convective ones is not as different as in previous studies. They are 5.8\% (15 stars) and 7.0\% (37 stars), respectively. It is clear from the figure that this is due to some of the new very-low-mass systems showing larger discrepancies than the long-ago well-known systems such as CM\,Dra, for~instance. This stresses the need to still increase the number of well-characterized DLEBs with late-M type~components.

\begin{figure}[H]
\includegraphics[width=12 cm]{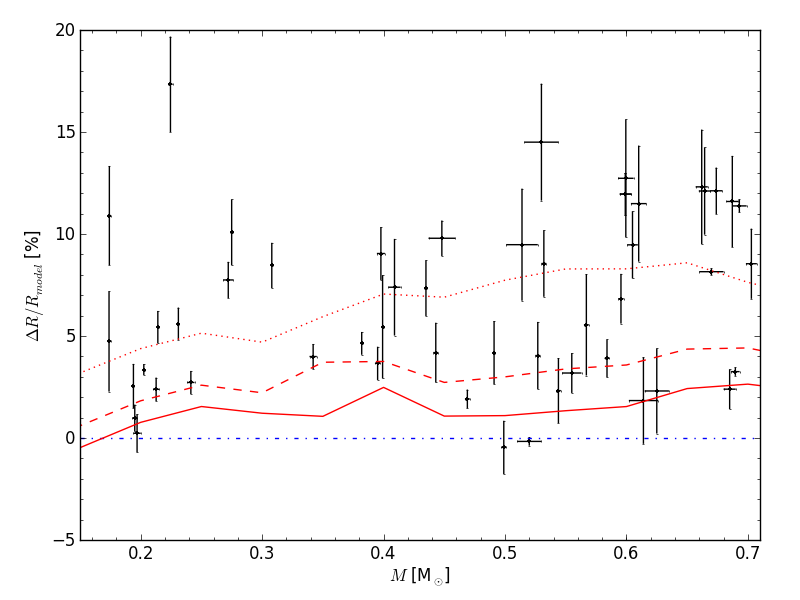}
\caption{Relative radius difference with respect to the BHAC15 1\,Ga stellar model as a function of the mass of the star. Red lines correspond to SPOTS models as labeled in Figure~\ref{fig:plot_M-R-Teff}. The~horizontal blue dot-dashed line is shown as a reference of agreement with the BHAC15 1\,Ga model. \label{fig:plot_relRdiff}.}
\end{figure}

As expected, the~discrepancies between the observations and models are reduced when the effect of photospheric spots is taken into account as in the case of the SPOTS theoretical models with $f > 0$. Stars are predicted to be larger and cooler, so that the mean radii discrepancies are reduced to 4.1\% and 0.5\% for models with filling factor 0.17 and 0.51, respectively. This is illustrated in Figure \ref{fig:plot_relRdiff}, where the relative difference between the measured radius of DLEB components and those predicted by the 1 Ga BHAC15 model ($\Delta R / R_{\rm model}$) is plotted. About 2/3 of the stars with $M \lesssim 0.7$\,M$_{\odot}$ in our sample could be explained assuming filling factors between 0 and 0.51; however, there are still few systems that show larger discrepancies, which may point to a larger effect of magnetic activity on their component stars. This makes crucial the implementation of such effects in stellar theoretical models as suggested in several works \citep{Chabrier2007,Morales2010,Feiden2013}.

\vspace{-6pt}

\section{Discussion}
\label{sec:discussion}
Different issues are still making it difficult to obtain a general picture of the problem between the observations of DLEBs and theoretical stellar structure models. \textls[-15]{For~instance, the~fundamental properties of stars depend on their metallicity and age. Both these properties are not as fundamentally constrained} as the masses and radii from light and radial velocity curves. Typically, the~age and metallicity are estimated from the membership of the system to moving groups or stellar clusters, or~in the case of metallicity from spectroscopic analysis. However, few of the low-mass stars with accurate masses and radii \textls[-15]{listed in Table~\ref{tab:sample} have reported values. This is the case of the classical systems CM\,Dra, whose age is estimated from the cooling} sequence of a white dwarf companion~\citep{Morales2009a}, YY\,Gem and CU\,Cnc, in~which cases their age is assumed from the membership to the Castor moving group \citep{Torres2002,Ribas2003}; however, assuming the corresponding ages and metallicities, the~radii discrepancies with respect to the Lyon stellar models are reduced by less than 1\% in the case of CM\,Dra and they are even larger for YY\,Gem and CU\,Cnc. This clearly demonstrates that stellar activity is playing a key role in the structure of such stars. The~case of NGTS\,0002-29 is also remarkable. The~system is a member of the Blanco\,1 open cluster \citep{Smith2021} with and estimated age of $90-150$\,Ma. Comparing with the BHAC15 models at this age range, radius discrepancies are reduced to $\sim$5.5\% and $\sim$3.2\% for components A and B, respectively. Still, there is a significant difference, although~in this case, it cannot yet be excluded that the system is still in the last stages of the pre-main sequence phase. On~the other hand, the~ages reported from isochrone fitting for the systems M55\,V54 \citep{Kaluzny2014} and UCA3\,127-192903 \citep{Kaluzny2013} show inconsistencies between the components of each binary. This points towards and additional effect causing the inflation of their~radii.

The main conclusion is that it is crucial to know the age and the composition of the DLEB and multiple systems in order to thoroughly test the stellar structure models. Efforts are on the way to characterize binary stars in clusters for that purpose. For~instance in a series of papers, Torres~et~al. analyzed several DLEBs pertaining to the Ruprecht\,147 open cluster \citep{Torres2018,Torres2019,Torres2020,Torres2021}. None of the systems fulfill the mass or accuracy threshold to be in Table~\ref{tab:sample}. However, it is worth mentioning that, interestingly, the~large mass components are generally well fitted by an isochrone at the expected age of the cluster, while the lower mass components still show radius discrepancies that can be attributed to the effect of stellar activity (see Torres~et~al. in this volume for more details \citep{Torres2021b}).

\textls[-10]{Throughout the paper we have also mentioned that there is broad evidence that stellar magnetic activity plays a crucial role in the stellar structure and evolution. As~stated in \mbox{Section \ref{sec:modelvsobs}}, several authors have taken its effects on stellar structure into account \mbox{\citep{Mullan2001,Chabrier2007,Feiden2013,Somers2020,Torres2021b}}} by implementing the impact of photospheric spots, of~strong magnetic fields, or~both in the theoretical models; however, this adds a new free parameter to the models that need to be tested against observations.
However, it is still uncertain how the stellar activity can be quantified and how it correlates with the radii inflation. DLEBs with active components typically show photometric variability caused by stellar spots with changing amplitude due to the evolution of spots. Frequently, they also show flaring activity on the light curves and also X-ray emission. Actually, L\'opez-Morales~et~al. \citep{LopezMorales2007a} obtained a linear relation between the X-ray to bolometric luminosity ratio ($L_{\rm X}/L_{\rm Bol}$) and the radius inflation of DLEBs ($\Delta R / R_{\rm model}$) that could be used as a correction to the observed radius values. Taking advantage of the second ROSAT all-sky survey \citep{ROSAT2} and the recent parallax determinations from Gaia DR3 \citep{Gaia2016,Gaia2022} we revisited such calibrations. Table\,\ref{tab:sampleLx} lists the binary systems with X-ray data. Bolometric luminosities were computed from the radius and effective temperature of each star. ROSAT data were converted to X-ray luminosities following the prescriptions in Schmitt~et~al. \citep{Schmitt1995} and the distance reported in the Gaia archive. We estimated the luminosity corresponding to each star by weighting by $v_{\rm sync}^{2}$ \citep{LopezMorales2007a}. This is equivalent to assuming a weighting according to $R^{2}$, i.e.,~that the X-ray luminosity of each component depends on the surface of the star, which can be used as a proxy for the surface of the~chromosphere.

\begin{table}[H]
\tablesize{\scriptsize}
\caption{List of DLEBs with X-ray data and parallax determination from~Gaia.}
\label{tab:sampleLx}
\begin{adjustwidth}{-\extralength}{0cm}
		\newcolumntype{C}{>{\centering\arraybackslash}X}
\newcolumntype{L}{>{\raggedright\arraybackslash}X}
		\begin{tabularx}{\fulllength}{LCCCCC}
\toprule
\textbf{Name}                         & \boldmath{$\varpi$}                & \textbf{X}                       & \textbf{HR}                         & \textbf{com.} & \boldmath{$L_{\rm X}/L_{\rm Bol}$} \\
                                      &    \textbf{[mas]}                         &\boldmath{ \textbf{[ct\,s}$^{-1}$\textbf{]}}                   &                                     &               &\boldmath{ \textbf{[}$\times$\textbf{10}$^{-4}$]}               \\
\midrule
\multirow{2}{*}{CM Dra}               & \multirow{2}{*}{67.288 $\pm$ 0.034}& \multirow{2}{*}{0.210 $\pm$ 0.017} & \multirow{2}{*}{$-$0.344 $\pm$ 0.061}  & A             & 8.9 $\pm$ 1.2  \\
                                      &                                  &                                  &                                     & B             & 9.0 $\pm$ 1.2  \\
\multirow{2}{*}{NGTS 0002-29 $^{1}$}   & \multirow{2}{*}{3.80 $\pm$ 0.26}   & \multirow{2}{*}{$\cdots$}        & \multirow{2}{*}{$\cdots$}           & A             & 14.8 $\pm$ 1.6 \\
                                      &                                  &                                  &                                     & B             & 17.6 $\pm$ 2.1 \\
\multirow{2}{*}{CU Cnc}               & \multirow{2}{*}{60.060 $\pm$ 0.036}& \multirow{2}{*}{0.732 $\pm$ 0.050} & \multirow{2}{*}{$-$0.084 $\pm$ 0.054}  & A             & 16.3 $\pm$ 4.1 \\
                                      &                                  &                                  &                                     & B             & 17.1 $\pm$ 4.6 \\
\multirow{2}{*}{NSVS01031772}         & \multirow{2}{*}{16.572 $\pm$ 0.018}& \multirow{2}{*}{0.076 $\pm$ 0.013} & \multirow{2}{*}{0.016 $\pm$ 0.17}     & A             & 7.0 $\pm$ 2.2  \\
                                      &                                  &                                  &                                     & B             & 8.2 $\pm$ 2.5  \\
YY Gem                                & 66.310 $\pm$ 0.023                 & 3.716 $\pm$ 0.091                  & $-$0.152 $\pm$ 0.020                   & A\&B          & 13.5 $\pm$ 1.7 \\
\multirow{2}{*}{GU Boo}               & \multirow{2}{*}{6.187 $\pm$ 0.011} & \multirow{2}{*}{0.039 $\pm$ 0.012} & \multirow{2}{*}{0.16 $\pm$ 0.32}      & A             & 17.3 $\pm$ 7.0 \\
                                      &                                  &                                  &                                     & B             & 19.4 $\pm$ 8.1 \\
\multirow{2}{*}{BD-15 2429}           & \multirow{2}{*}{23.58 $\pm$ 0.14}  & \multirow{2}{*}{0.495 $\pm$ 0.042} & \multirow{2}{*}{$-$0.249 $\pm$ 0.068}  & A             & 7.0 $\pm$ 1.7  \\
                                      &                                  &                                  &                                     & B             & 8.0 $\pm$ 2.3  \\
\multirow{2}{*}{RXJ0239.1-1028}       & \multirow{2}{*}{8.144 $\pm$ 0.016} & \multirow{2}{*}{0.036 $\pm$ 0.014} & \multirow{2}{*}{1.00 $\pm$ 0.59}      & A             & 5.2 $\pm$ 2.4  \\
                                      &                                  &                                  &                                     & B             & 7.3 $\pm$ 3.3  \\
\multirow{2}{*}{EPIC 247605441}       & \multirow{2}{*}{9.99 $\pm$ 0.30} & \multirow{2}{*}{0.129 $\pm$ 0.019} & \multirow{2}{*}{$-$0.06 $\pm$ 0.10}      & A             & 2.7 $\pm$ 0.5  \\
                                      &                                  &                                  &                                     & B             & 16.7 $\pm$ 3.8  \\                                      
\multirow{2}{*}{ASAS J065134-2111.5}  & \multirow{2}{*}{10.819 $\pm$ 0.021}& \multirow{2}{*}{0.049 $\pm$ 0.012} & \multirow{2}{*}{$-$0.35 $\pm$ 0.21}    & A             & 0.69 $\pm$ 0.22\\
                                      &                                  &                                  &                                     & B             & 2.56 $\pm$ 0.86\\
\multirow{2}{*}{IM Vir}               & \multirow{2}{*}{11.136 $\pm$ 0.017}& \multirow{2}{*}{0.261 $\pm$ 0.031} & \multirow{2}{*}{$-$0.232 $\pm$ 0.089}  & A             & 3.38 $\pm$ 0.56\\
                                      &                                  &                                  &                                     & B             & 10.0 $\pm$ 2.0 \\
\multirow{2}{*}{V530 Ori}             & \multirow{2}{*}{9.763 $\pm$ 0.018} &\multirow{2}{*}{0.0252 $\pm$ 0.0092}& \multirow{2}{*}{$-$0.41 $\pm$ 0.26}    & A             & 0.36 $\pm$ 0.17\\
                                      &                                  &                                  &                                     & B             & 1.90 $\pm$ 0.94\\
\bottomrule
\end{tabularx}
	\end{adjustwidth}

\footnotesize{\textbf{$^{1}$} $L_{\rm X}/L_{\rm Bol}$ is estimated assuming $\log L_{\rm X} = 29.29$\,erg\,s$^{-1}$ \citep{Smith2021}.}
\end{table}

\begin{figure}[H]
\includegraphics[width=12 cm]{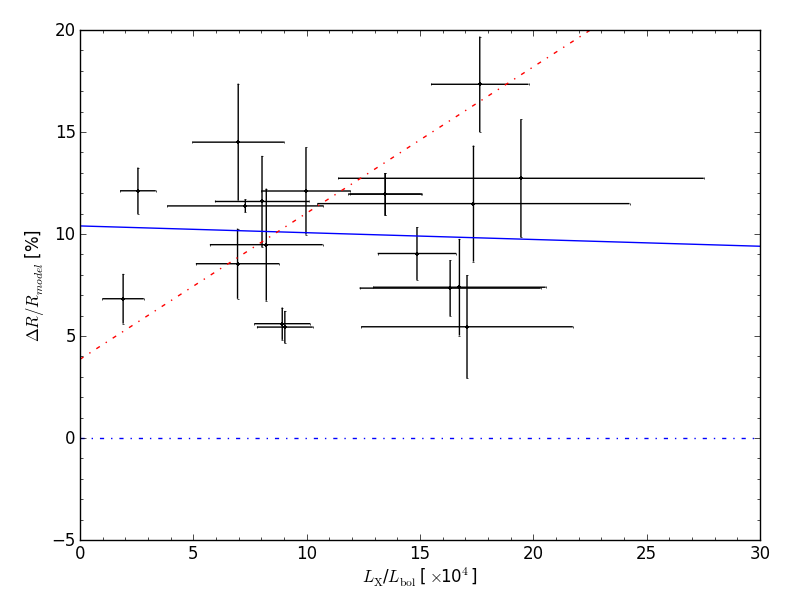}
\caption{Relative radius difference with respect to the BHAC15 stellar models at 1\,Ga as a function of the X-ray to bolometric luminosity ratio. The~solid-blue line depicts the best linear fit to the data, whose slope is not significantly different from zero. The red dot-dashed line illustrates the slope reported in L\'opez-Morales~et~al. \citep{LopezMorales2007a} fitted to our data for comparison. The~horizontal blue dot-dashed line is shown as a reference of agreement with the BHAC15 1\,Ga~model. \label{fig:plot_LxLbol}} 
\end{figure}

The comparison between $\Delta R/R_{\rm model}$ and $L_{\rm X}/L_{\rm Bol}$ is illustrated in Figure \ref{fig:plot_LxLbol}. $\Delta R/R_{\rm model}$ is again computed taking the 1\,Ga BHAC15 model as reference. The~correlation factor of these dataset, $\rho$ = 0.12 (\emph{p}-value = 0.64), points towards uncorrelation. Actually, the~slope of the best fit (blue solid line) is not significantly different from zero. From~this result, we conclude that a linear relation is not supported by our sample of late-type dwarfs in DLEBs. We stress here that this may
be caused by the fact that the low-mass stars in our sample show saturated levels of X-ray emission. Indeed, the~X-ray to bolometric luminosity ratio of the stars illustrated in this figure is in the range   $-3.6 < \log (L_{\rm X}/L_{\rm Bol}) < -2.7$, well in the saturated regime. The~updated X-ray data and distances are also responsible for some of the differences with the previous work \citep{LopezMorales2007b}. For~instance, the~distance to the CU\,Cnc system provided by Gaia is about 20\% farther than previously reported \citep{Ribas2003}, therefore its X-ray emission is~stronger.

\vspace{-6pt}

\begin{figure}[H]
\includegraphics[width=11 cm]{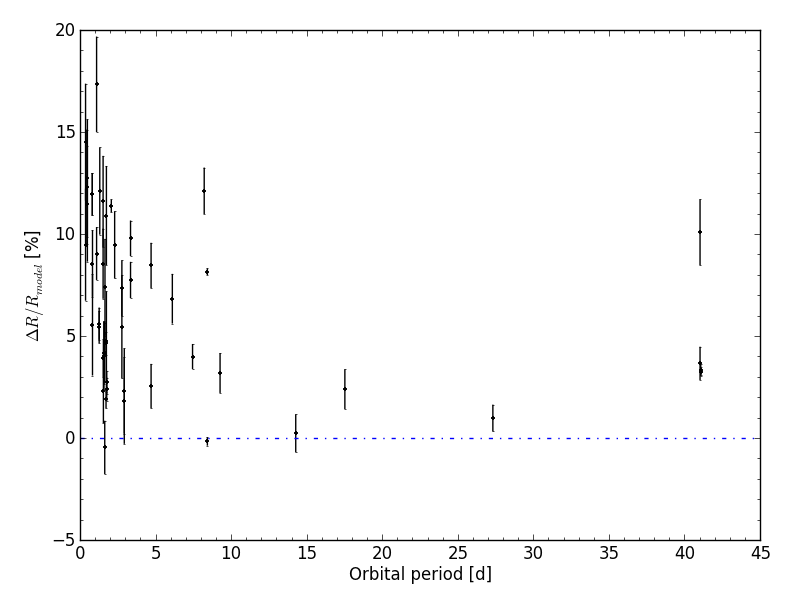}
\includegraphics[width=11 cm]{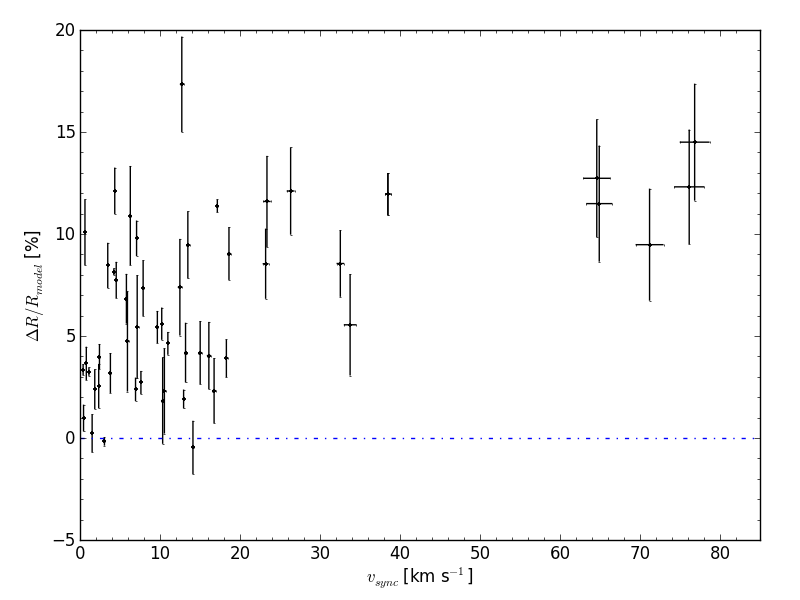}
\caption{Relative radius difference with respect to the BHAC15 stellar models at 1\,Ga as a function of the orbital period of the system (\textbf{top}) and the pseudo-synchronization rotation velocity at periastron (\textbf{bottom}). The~horizontal blue dot-dashed line is shown as a reference of agreement with the BHAC15 1\,Ga~model. \label{fig:plot_vrot}}
\end{figure}

Certainly, there is a well known correlation between X-ray emission (and activity level) and the rotation of stars \citep{Pizzolato2003} suggesting that the emission of late-type stars is saturated for rotation periods $\lesssim$10\,days. This may be the case for most of the systems in Table\,\ref{tab:sample}, whose orbital periods are below few days and the rotation of their components may be synchronized to the orbital motion as mentioned in the previous section. DLEBs with long period orbits, whose components may not be rotating synchronously, may be less magnetically active and more consistent with models. This was statistically studied by Coughlin~et~al. \citep{Coughlin2011} using eclipsing binaries detected in the \textit{Kepler} mission field-of-view. From the analysis of only their light curves, they concluded that the mean radius discrepancy between the DLEBs measurements and theoretical models diminishes for stars in systems with longer orbital periods. For~comparison, the~panels in Figure \ref{fig:plot_vrot} illustrate the $\Delta R/R_{\rm model}$ with respect to the orbital period of the system and $v_{\rm sync}$. It is evident from these plots that the DLEBs with $v_{\rm sync}$ above $\sim 20$\,km\,s$^{-1}$ show larger discrepancies on average, although they are a few. On the contrary, the radii of DLEBs components with $v_{\rm sync} \lesssim 4$\,km\,s$^{-1}$ and orbital periods above $\sim$8 days are closer to model predictions. This reinforces the hypothesis of stellar activity, caused by fast stellar rotation, as~playing a crucial role in the structure of~stars.

The DLEBs with long orbital periods discovered in the past decade deserve some attention. For~instance, Kepler-16 \citep{Doyle2011} and LSPM J1112+7626 \citep{Irwin2011}, both have an orbital period of $\sim$41 days, the~longest in our sample.
Although this long orbital period, $\Delta R/R_{\rm model}$ is about 3.25\% and 3.35\% for Kepler-16 A and B components, respectively, and~3.7\% and  10.1\% for LSPM J1112+7626 A and B, respectively. However, their light curves still show 1\% and 2\% out-of-eclipse photometric variability, respectively. The period of such variability is consistent with spin-orbit pseudo-synchronization for Kepler-16, while it is about twice larger for LSPM J1112+7626, which points towards slowly rotating components; however, the~photometric variability suggests a moderate activity level. On the contrary, Kepler-453, with~a shorter orbital period of $\sim$27.3 days, shows a lower level of stellar activity and a low-mass component fairly in good agreement with a 1\,Ga theoretical model. The~only caveat is a slightly different radii ratio than predicted by models \citep{Welsh2015}. This is also the case of KIC 9821078 ($P$ = 8.9 days). As~illustrated in the top panel of Figure \ref{fig:plot_vrot}, the~less massive component is well fitted by the 1\,Ga BHAC15 theoretical model, however, a~$\Delta R/R_{\rm model} \sim$8\% is found for component A. The~same happens for the shorter period system MG1 646680 ($P$ = 1.64 days), with~the A component well reproduced by models but B showing a $\sim$4\% inflated radii. Although~this is a well-known issue when fitting light curves of DLEBs showing only partial eclipses, in~which case the radii and luminosity ratios are correlated, the~sum of the radii is better constrained; therefore, this points towards and inflated radii for both components rather than to inconsistent ages between~them.

EBLM J0113+31 is also an interesting binary with a relatively long orbital period of 14.3 days. The~system, composed by a solar-type star and an late-M dwarf, was first reported as a single-lined single-eclipse binary system \citep{GomezMaqueoChew2014}. It was later observed from space with \textit{TESS} \citep{TESS} and \textit{CHEOPS} \citep{CHEOPS2021} unveiling the eclipses of the fainter component, and~its radial velocity semi-amplitude was retrieved as well from high-resolution spectroscopic observations \citep{Swayne2020,Maxted2022}. Thus, precise masses and radii are fundamentally determined for both components. In~contrast to the systems reported above, both components of this system are well reproduced by stellar models with an age of 6.7\,Ga \citep{Maxted2022}. Curiously, \textit{TESS} photometry does not show any variability signature. Similarly, KIC 6131659 is also formed by a solar-type star and an early M dwarf orbiting around each other every $\sim$17.5 days. The~photometric variability of the system is well below the 1\% level, which indicates low stellar activity. Both components are in also in good agreement with theoretical models at an age of 3.5\,Ga \citep{Bass2012}.
Interestingly, IM\,Vir is also a system with very similar components in mass to KIC 6131659, but~with a much shorter period (1.31 days), larger photometric variability, and~X-ray emission, thus more magnetically active. In~this case, the~components cannot be fitted to the same theoretical isochrone and they actually show inflated radii \citep{Morales2009b}. The~same issue is also found for the similar DLEB ASAS J065134-2111.5 composed of a solar-type star and a late K-type star at an intermediate orbital period of 8.22 days, and~with high levels of magnetic activity as well (X-ray and Ca H\&K emission \citep{Helminiak2019}); and also for the system EPIC 247605441 composed by a solar type star and an M dwarf orbiting at an orbital period of 2.29 days \citep{Helminiak2021}. The~contrast between these different systems makes evident the strong impact of magnetic activity on the stellar properties and complicates the global comparison of the observed radii with theoretical stellar structure and evolution models. The discovery and analysis of further long period systems will help to study the properties of stars with different magnetic activity levels.

\section{Conclusions}
\label{sec:conclusions}
The studies of low-mass stars in DLEBs appeared in the past decade have now made evident that the problem of the radius inflation of late-type stars is caused by their intrinsic high level of magnetic activity that is manifested as surface spots, chromospheric emission, strong X-ray luminosity, and~flaring activity. As~reported throughout this work, new DLEBs and multiple systems showing different levels of stellar activity and theoretical works support this hypothesis. This adds an additional difficulty when testing stellar structure and evolution models because a new parameter comes into play, stellar activity. Even if this is taken into account as a surface filling factor, the~strength of the interior magnetic field, the~flux of emission lines, or~high-energy radiation, further observations are needed to constraint these quantities. For~this purpose, it would be useful to estimate the activity level in DLEB studies uniformly. In~this context, new all-sky X-ray surveys such as eROSITA \citep{eROSITA} may help to re-evaluate again the correlation between the radius discrepancy of M dwarfs and high-energy emission for non-saturated~systems.

More DLEB and multiple systems with accurate fundamental properties as well as updated models including stellar activity will help to finally quantify the effect of activity on the structure of stars and to test theoretical predictions. Better determinations of the effective temperature of DLEBs components would also help here by adding an additional reliable constraint. The~extremely precise distances that Gaia is providing can help on that. Individual temperatures can be estimated from the temperature ratio derived from the light curve analysis and the absolute luminosity of the system making use of bolometric~corrections.

This work also reveals that further DLEBs with extremely accurate masses and radii are still needed. On~the one hand, only a handful of very-late type stars with accuracy below the 3\% level are known. New systems are showing that radii difference with respect to theoretical models may be larger than previously found. On~the other hand, there are just a few systems in the region around $\sim$0.35\,M$_{\odot}$ and the~boundary between fully and partially convective stars. More observations will allow us to better understand this transition region. In~the case of early type stars, although~the number of known systems is significantly larger, the~interpretation is hampered by the dispersion also caused by age. For~this purpose, DLEB in stellar clusters, from~which the age and metallicity can be derived have been proven to be helpful \citep{Torres2021b}.

It is also evident that binary stars with long orbital periods are particularly useful to understanding the radius inflation problem. In~this work, we have discussed several systems that show different levels of radii discrepancies depending on their activity properties, some of them showing agreement with theoretical models. Additional systems with different levels of stellar activity will help to constraint its effects on the radii of stars. Although~these DLEBs are more difficult to find, because~of the lower probability of being eclipsing and the scarcity of eclipses, photometric surveys such as \textit{Kepler} or \textit{TESS} succeeded in finding several of them, and~more may come from missions such as \textit{PLATO} \citep{PLATO2014} and ground-based exoplanet surveys. Efforts to accurately characterize their properties are worthwhile to better understand the properties of low-mass stars and the evolution of close binary systems. A~better knowledge of such type of stars will also benefit the determination of the properties of the exoplanets they host, which are nowadays prime targets to look for Earth-like planets and to study their~exo-atmospheres.

\vspace{6pt} 


\acknowledgments{This publication has been made possible by grants PGC2018-098153-B-C33 funded by 
MCIN/AEI/10.13039/501100011033 and by “ERDF A way of making Europe”. We acknowledge as well the support of the Generalitat de Catalunya/CERCA programme. This work was also partially supported by the program Unidad de Excelencia María de Maeztu~CEX2020-001058-M. This work has made use of data from the European Space Agency (ESA) mission Gaia (\url{https://www. cosmos.esa.int/gaia}
 processed by the Gaia Data Processing and Analysis Consortium (DPAC, \url{https://www.cosmos.esa.int/web/gaia/dpac/ consortium}. Funding for the DPAC has been provided by national institutions, in~particular the institutions participating in the Gaia Multilateral Agreement.}

\begin{adjustwidth}{-\extralength}{0cm}
\reftitle{References}

\externalbibliography{yes}

\end{adjustwidth}
\end{document}